\documentclass[aps,prl,twocolumn, superscriptaddress]{revtex4-2}
\usepackage{graphicx}
\usepackage{amsmath}
\usepackage{hyperref}
\usepackage{textcomp}

\usepackage[T1,T2A]{fontenc}
\usepackage[utf8]{inputenc}
\usepackage[russian,english]{babel}
\usepackage[dvipsnames]{xcolor}

\begin{document}
\title{
Probing intermittent polariton vortex dynamics with two-point correlations}
\author{R.\,Cherbunin}
\email{r.cherbunin@spbu.ru}
\affiliation{Spin Optics Laboratory, Saint-Petersburg State University, St. Petersburg, 198504, Russia}
\affiliation{Russian Quantum Center, Skolkovo, Moscow, 121205, Russia}
\author{A.\,Liubomirov}
\affiliation{Russian Quantum Center, Skolkovo, Moscow, 121205, Russia}
\author{P.\,Kozhevin}
\affiliation{Spin Optics Laboratory, Saint-Petersburg State University, St. Petersburg, 198504, Russia}
\affiliation{Abrikosov Center for Theoretical Physics, MIPT, Dolgoprudnyi 141701, Russia}
\author{M.\,Chukeev}
\affiliation{Spin Optics Laboratory, Saint-Petersburg State University, St. Petersburg, 198504, Russia}
\author{M.\,Petrov}
\affiliation{Spin Optics Laboratory, Saint-Petersburg State University, St. Petersburg, 198504, Russia}
\affiliation{Russian Quantum Center, Skolkovo, Moscow, 121205, Russia}
\author{S.\,Kavokina}
\affiliation{Spin Optics Laboratory, Saint-Petersburg State University, St. Petersburg, 198504, Russia}
\affiliation{Russian Quantum Center, Skolkovo, Moscow, 121205, Russia}
\affiliation{Abrikosov Center for Theoretical Physics, MIPT, Dolgoprudnyi 141701, Russia}

\author{A.\,Nalitov}
\email{anton.nalitov@gmail.com}
\affiliation{Russian Quantum Center, Skolkovo, Moscow, 121205, Russia}
\affiliation{Abrikosov Center for Theoretical Physics, MIPT, Dolgoprudnyi 141701, Russia}

\author{A.\,Kavokin}
\email{kavokinalexey@gmail.com}
\affiliation{Spin Optics Laboratory, Saint-Petersburg State University, St. Petersburg, 198504, Russia}
\affiliation{Russian Quantum Center, Skolkovo, Moscow, 121205, Russia}
\affiliation{Abrikosov Center for Theoretical Physics, MIPT, Dolgoprudnyi 141701, Russia}
\affiliation{School of Science, Westlake University, Hangzhou 310024, China}

\date{\today}

\begin{abstract}
Superconducting single-photon detectors with time resolution of 80 ps have been used to study the spatiotemporal dynamics of a trapped bosonic condensate of exciton-polaritons.
Both second- and first-order correlation functions are found to exhibit pronounced oscillations in time governed by the dynamics of the polariton condensate in the trap.
We have identified the intermittent regime of stochastic transitions between stationary and limit-cycle regimes near the Andronov-Hopf bifurcation manifested in asymmetric distortions of the correlation function.
This rich interplay of deterministic and stochastic condensate dynamics is explained theoretically as a manifestation of mutually compensating self-repulsion and reservoir-mediated attraction.
\end{abstract}

\maketitle

Interaction-induced correlations lie at the heart of many-body physics and present the underpinning mechanisms of Bose-Einstein condensation, superconductivity, and lasing \cite{pethick2008,leggett2006,scully1997}.
In ultracold atomic gases, second-order correlations wese used to probe the BEC-BCS crossover, revealing the transition from fermionic pairing to molecular condensation through statistics of density fluctuations \cite{giorgini2008,bloch2008}.
In quantum optics, the Glauber correlation functions together with higher-order correlations, enable the reconstruction of the full quantum state beyond the mean-field level \cite{glauber1963,glauber1963a,loudon2000}.

In the context of exciton-polaritons, spatiotemporal correlations play a pivotal role and allow unveiling the many-body physics of polariton phases unfolding inside microcavities through the statistics of their optical emission \cite{carusotto2013,byrnes2014}.
The second-order correlation function $g^{(2)}(\tau)$ was crucial in demonstrating the transition from a thermal state to a coherent polariton condensate \cite{kasprzak2006}.
In the out-of-equilibrium regime of polariton lasing, spontaneous coherence buildup enables the probing of spatiotemporal phase and density profiles via interferometric optical measurements \cite{alyatkin2020,Barrat2024}.
Moreover, independent measurements of the first- and second-order correlation functions allow one to separate the phase and density coherence times, both of which can significantly exceed the polariton lifetime \cite{Love2008,Adiyatullin2015}.
Recent discoveries of emergent supersolidity \cite{Trypogeorgos2025,kozhevin2025} and universal Kardar–Parisi–Zhang scaling \cite{Fontaine2022,Widmann2026} in nonequilibrium polariton condensates, enabled by correlation function measurements, further underscore the potentiality of these probes for unraveling novel exotic phases of hybrid light–matter systems.

Polariton condensates trapped in optically tailored effective potential wells are characterised by extremely long coherence times due to spatial separation of the condensate from the main source of its decoherence, which is the exciton reservoir \cite{Demenev2016,Orfanakis2021}.
The latter simultaneously plays a dual role: it compensates spontaneous polariton radiative losses in the condensate via bosonic stimulated scattering and provides an effective trapping potential due to repulsive exciton exchange interactions \cite{Askitopoulos2013,Askitopoulos2015}.
As the resulting effective potential is complex-valued, this two-component system presents a peculiar platform for implementing non-Hermitian quantum mechanics \cite{Gao2015,Gao2018}.
At the same time, the interplay of intra- and intercomponent linear and nonlinear couplings, and significantly different characteristic timescales of the condensate and reservoir, renders the conventional complex Ginzburg-Landau description insufficient for correct stability analysis \cite{Chestnov2024}.
In contrast, the extended Gross-Pitaevskii model beyond the adiabatic approximation, which eliminates the reservoir dynamics, has been used to predict nonequilibium supersolidity and to reproduce vorticity oscillations in optically trapped polariton condensates \cite{Nalitov2019,Barrat2024,kozhevin2025}.

In this Letter, we investigate the two-point second-order correlation function $g^{(2)}(\tau,\mathbf{r}_1,\mathbf{r}_2)$ of an optically trapped polariton condensate in a dynamical superposition of two counter-rotating vortex modes.
The observed oscillations of $g^{(2)}$ at the frequency of the spectral splitting in the condensate emission indicate persistent periodic condensate dynamics related to the trap ellipticity.
We reproduce this oscillatory regime using the extended Gross–Pitaevskii equation (eGPE) by projecting the dynamics onto the superposition ansatz of two counter-rotating vortex states with topological charges $l=\pm1$.
Within this framework, adopted from Ref. \cite{Chestnov2024}, the condensate density oscillations can be associated to Larmor precession of the condensate orbital angular momentum (OAM) pseudospin around an effective field induced by weak trap ellipticity.
At lower pumping powers, the observed asymmetric distortions of the correlation function reflect intermittent transitions between stationary and self-oscillatory regimes, separated by a supercritical Andronov-Hopf (AH) bifurcation and induced by slow fluctuation of the pump power.

\begin{figure}[htbp]
    \centering
    \includegraphics[width=\columnwidth]{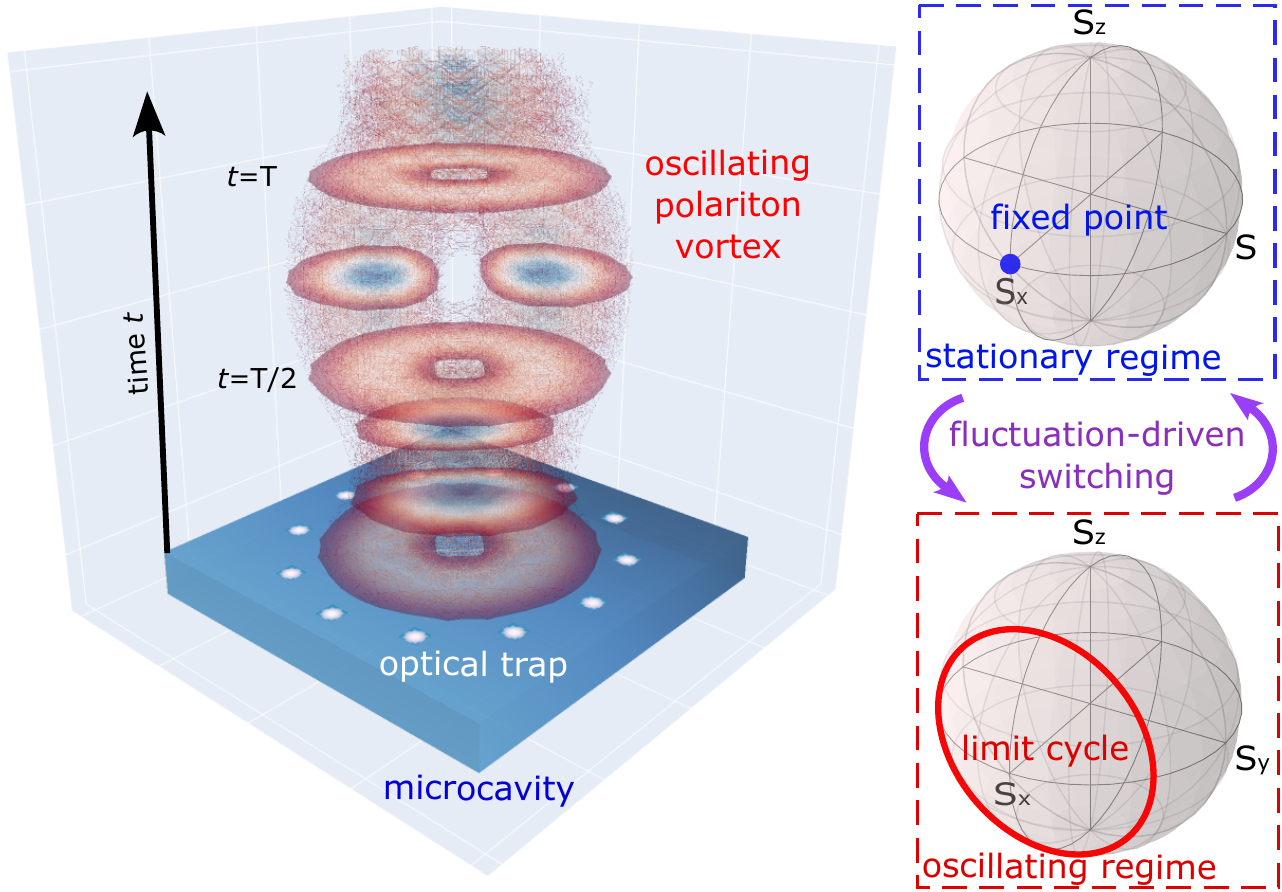}
    \caption{Sketch of intermittent polariton condensate dynamics  near a bifurcation point: fluctuations induce transitions between the stationary (fixed-point) and oscillatory (limit-cycle) regimes. The latter is manifested in spatiotemporal polariton vortex dynamics (left panel).}
    \label{fig:one}
\end{figure}


The experiments were carried out on a high-Q microcavity with Bragg mirrors made of gallium arsenide and aluminum arsenide, similar to that used in \cite{opt6040053}, at a temperature of 4 K.  The optical detuning of the photonic mode was chosen to be slightly negative, corresponding to a large polariton diffusion length, several times exceeding the trap diameter. An optical trap with a size of about 12 microns and eppilticity about 5\% was created on the sample using a single-mode laser, whose spatial profile was shaped by a spatial light phase modulator. The emission was detected in a reflection geometry through a dichroic mirror in a single linear polarization. The measurement of the first-order coherence time was performed using a Mach-Zehnder interferometer. For this purpose, the condensate emission was split into two beams, after which one beam passed through a 1 m long motorized delay line. Then both images, having experienced the same number of reflections, were recombined on a beam-splitting cube and detected by a cooled CCD camera. To measure the second-order correlation function, the emission was split into two channels using a beam-splitting cube, and each of the beams was focused by a short-focus lens onto an adjustable pinhole with a diameter of 75 $\mu$m. The position of the condensate image on the pinhole was monitored by back-reflection using a CCD camera. The two pinholes allowed for the independent selection of emission from two points of the condensate, thereby measuring the intensity correlation function at different spatial points. The emission that passed through the pinholes was focused by two micro-objectives into single-mode polarization-maintaining optical fibers, after which it was detected by superconducting detectors with a time resolution of 80 ps. To measure the photon-counting statistics, a time-interval counting unit with a resolution of 8 ps was used.


The study of the first- and second-order correlation function was carried out at four points at a fixed distance $R\approx 5$ $\mu$m from the trap center, where the radial condensate density peaks, for four azimuthal angles relative to the horizontal axis:  $\varphi_X = 0$, $\varphi_Y = \pi/2$, $\varphi_A = \pi/4$, and $\varphi_D = -\pi/4$ (see the map of the condensate with indicated probe points in Figure 2(d)).
Measured emission intensities $I(t,\mathbf{r}_j)$ at the four points $j=\text{X,Y,A,D}$ were processed using the conventional definition of the two-point correlation function:
\begin{equation} \label{eq:g2}
    g^{(2)}_{j,k}(\tau,\mathbf{r}_j,\mathbf{r}_k) = \frac{\langle I(t,\mathbf{r}_j) I(t+\tau,\mathbf{r}_k) \rangle_t}{\langle I(t,\mathbf{r}_j) \rangle_t\langle I(t,\mathbf{r}_k) \rangle_t}.
\end{equation}

\begin{figure*}
    \centering
    \includegraphics[width=2\columnwidth]{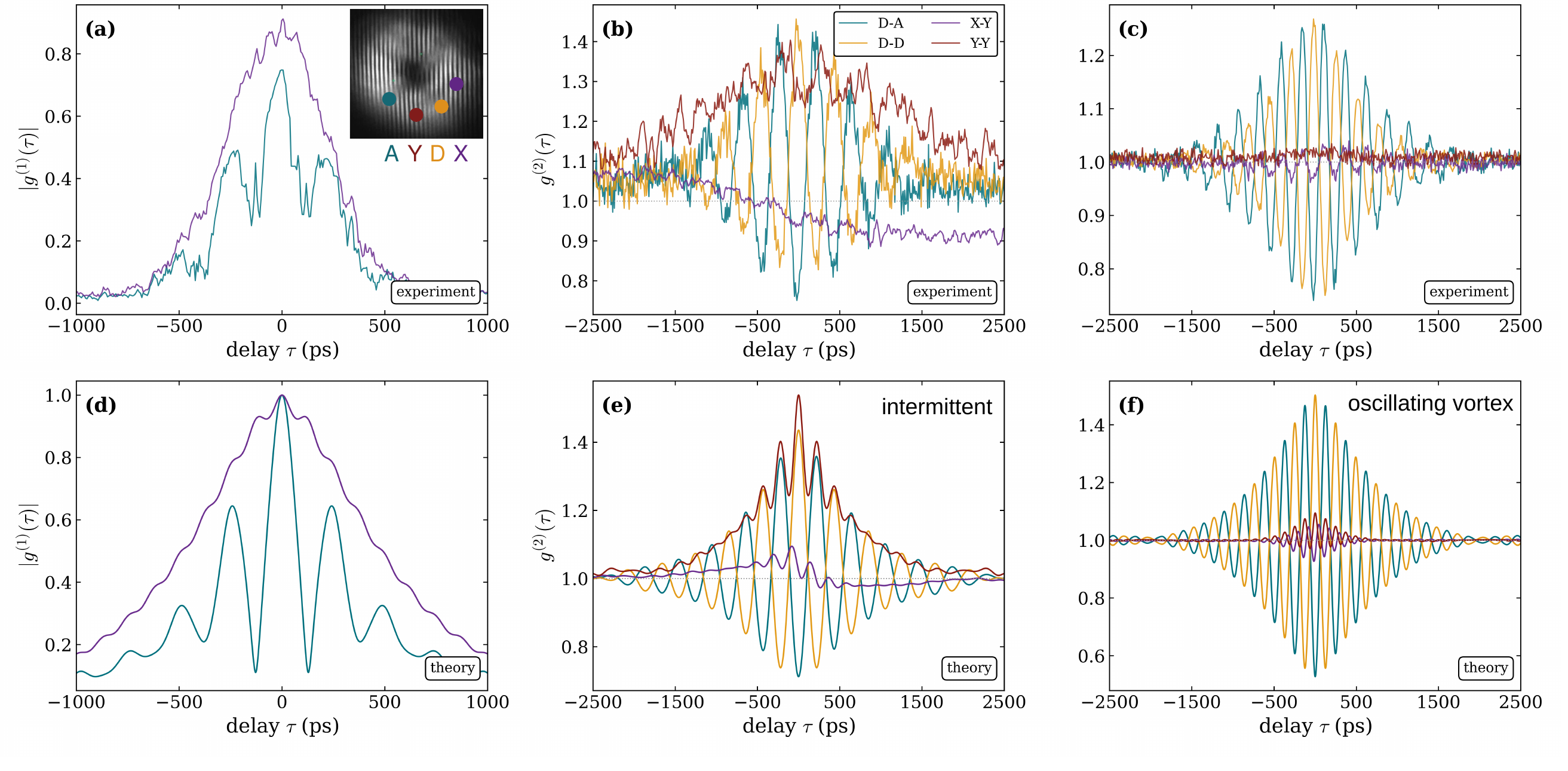}
    \caption{(Color online) Two-point correlation function measurements.
    Left column: reference first-order correlation function at point A (blue) and X (purple), experiment (a) and theory (d).
    Central column: second-order correlation function in D-A (blue), D-D (yellow), Y-Y (red), and X-Y (blue) configurations, experiment (b) and theory (e).
    Right column: same as central column, but for the oscillating polariton vortex regime at higher pumping power, in experiment (c) and in theory (f). The inset in Figure 2(a) shows an example of an interferogram of the trapped polariton condensate with indicated locations of four probes: X,Y, A, D.}
    \label{fig:two}
\end{figure*}

The results of the two-point second-order correlation function are presented in Figure \ref{fig:two} for two qualitatively different regimes observed at two pumping powers.
In both cases, shown in Figs.\ref{fig:two}b,c, pronounced oscillations decaying to unity with a characteristic time  $\tau$  are observed in DA and DD configurations with periodicity corresponding to previously observed condensate oscillations  \cite{opt6040053}.
At low pump power, the oscillations are strongly asymmetric and range from $\sim$0.8 to $\sim$1.4, in stark contrast to symmetric oscillations at high pump power.
The most dramatic difference between the two regimes, however, is observed in the XY and YY two-point configurations, corresponding to the two main axes of the elliptic trap.
While at high pump power both correlation functions barely deviate from unity, in the low-pump-power case the YY correlations reach the same values as those of DD, DA and appear to follow their envelope at long timescales.

We describe the polariton condensate wavefunction in a superposition of two quasi-degenerate eigenstates in a weakly elliptic optically induced trapping potential $\Psi_x \propto \cos(l\varphi)$ and $\Psi_y \propto \sin(l\varphi)$ with $l=\pm1$.
It is convenient to use the basis of vortex states $\Psi_\pm = (\Psi_x\pm\Psi_y)/\sqrt{2}$, in which the superposition state $\psi_+\Psi_++\psi_-\Psi_-$ can be described with a pseudovector $\mathbf{S}=\psi^\dagger \boldsymbol{\sigma}\psi$, where $\psi = (\psi_+,\psi_-)^\mathrm{T}$ \cite{Askitopoulos2018}.
Beyond the conventionally employed adiabatic approximation, its dynamics is governed by the vector equation \cite{Chestnov2024}
\begin{equation} \label{eq:s_dyn}
    \dot{\mathbf{S}} = N_0 \mathbf{S} + S\mathbf{N} - \left[ (\varepsilon\mathbf{N}+\xi\mathbf{S}) \times \mathbf{S}_\perp \right],
\end{equation}
where $\varepsilon$ and $\xi$ parameters govern the condensate self-interaction and interaction with the trap, respectively, while $N_0$ and $\mathbf{N}$ describe the trap density, whose dynamics is in turn coupled to that of the condensate:
\begin{subequations} \label{eq:resdyn}
\begin{eqnarray}
    \dot{N_0} =& W - (1+2S)N_0 -(SN_\text{th}+\mathbf{S}\cdot\mathbf{N}), \label{eq:n0_dyn} \\
    \dot{\mathbf{N}} =& \delta \mathbf{W} - (1+2S)\mathbf{N} - 2\mathbf{S}_{||}(N_\textrm{th}+N_0),  \label{eq:n_dyn}
\end{eqnarray}
\end{subequations}
where $N_\mathrm{th}$ is the ratio of the trap exciton and condensate polariton lifetimes, while $W$ and $\delta \mathbf{W}=(\delta W,0)$ correspond to the azimuthally uniform and angular-dependent parts of the pumping power.
The out-of-plane (z) and in-plane (x-y) projections of pseudovector $\mathbf{S}$ are denoted as $\mathbf{S}_\perp$ and $\mathbf{S}_{||}$, respectively.


\begin{figure}
    \centering
    \includegraphics[width=1.0\linewidth]{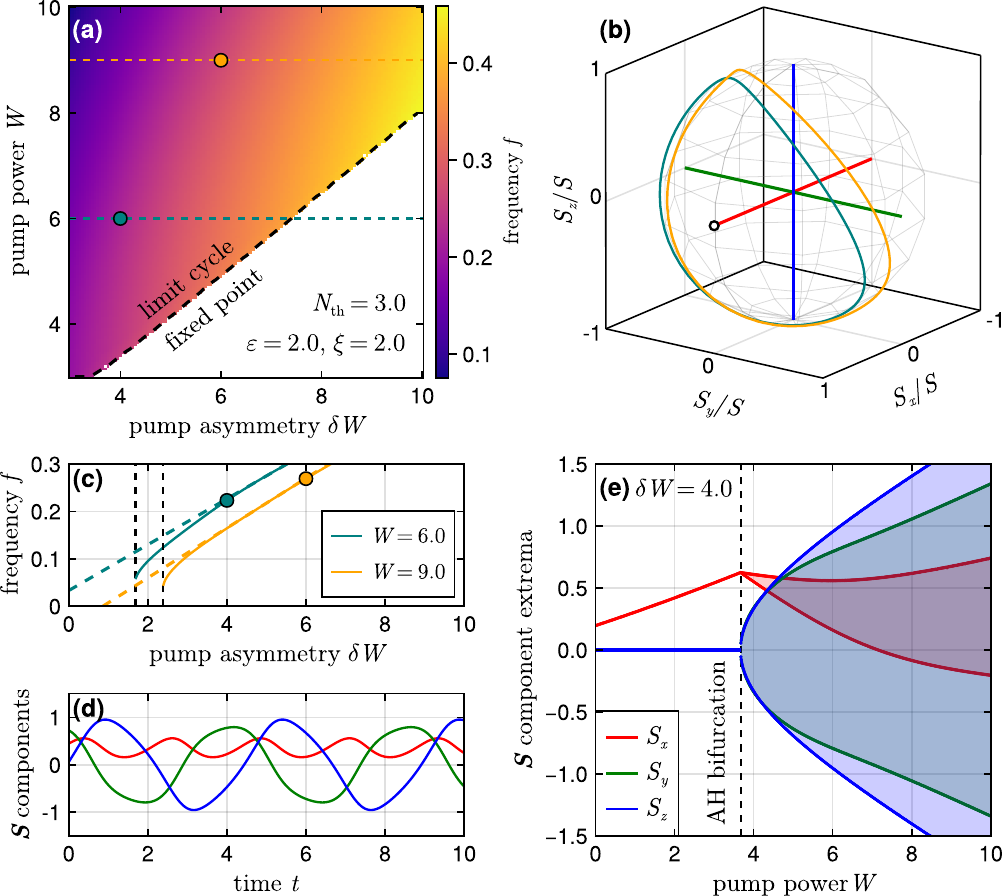}
    \caption{Deterministic condensate limit-cycle dynamics in the eGPE model (\ref{eq:s_dyn},\ref{eq:resdyn}).
    (a) Frequency map on the $(\delta W,W)$ plane.
    (b) Limit cycles of the pseudospin $\mathbf{s} = \mathbf{S}/S$ on the Bloch sphere for two points in panel (a). White marker shows the fixed point position for the stationary regime (white area in panel (a)).
    (c) Frequency cross-sections and linear fitting (dot-dashed lines) for two pump powers $W=6,9$ (dashed lines in panel (a)). Vertical dashed lines indicate AH bifurcation.
    (d) Periodic time dependence of the pseudovector components $S_x,S_y,S_z$ for the real limit cycle in panel (b).
    (e) AH bifurcation diagram: stationary values and variation ranges for the pseudovector components.
    Parameters used: $N_\text{th} = 3.0, \varepsilon = \xi = 2.0$.}
    \label{fig:three}
\end{figure}

In the framework of Eqs. (\ref{eq:s_dyn},\ref{eq:resdyn}), periodic condensate dynamics emerges in the self-oscillatory regime of stable limit cycles, illustrated in Fig.\ref{fig:three}.
This regime is present if the condensate repulsion and the reservoir-mediated attraction, governed by parameters $\xi,\varepsilon$, partly compensate each other \cite{Nalitov2019}.
Moreover, the onset of this regime is above the condensation threshold at a critical point, which scales with the degree of pump asymmetry and separates the self-oscillatory regime from the stationary condensate mode, as shown in Fig. \ref{fig:three}a.
Above the critical line in the plane $(\delta W,W)$, the oscillation frequency slowly descreases with the pump power $W$ and exhibits strong dependence on the ellipticity-induced pump asymmetry $\delta W$, which saturates to linear increment, as shown in Figs.\ref{fig:three}a,c.
The limit cycles are conveniently illustrated as closed paths of the pseudospin $\mathbf{s} = \mathbf{S}/S$ on the Bloch sphere, while the stationary condensate regime corresponds to a trivial fixed-point attractor $s_x=1$, $s_y=s_z=0$, as illustrated in Fig.\ref{fig:three}b.
The time dependence of the pseudovector components exhibits weak anharmonicity due to admixture of the second frequency harmonic, as shown in Fig. \ref{fig:three}d.
The onset of the self-oscillating regime occurs at a supercritical AH bifurcation corresponding to soft instability of a trivial fixed-point attractor and emergence of a stable limit cycle, as illustrated in Fig. \ref{fig:three}e.


Assuming proportionality of the condensate density to the normalized dimensionless emission intensity $|\Psi(\mathbf{r},t)|^2\propto I(\mathbf{r},t)$, the latter can be expressed in the components of the pseudovector $\mathbf{S}(t)$ using the identity $\psi_\pm =\sqrt{(S\pm S_z)}\exp(i\phi\pm i\delta\phi/2)$ with $\delta \phi = \arg(S_x+iS_y)$ and $\phi$ the global gauge phase of the condensate.
For the four experimentally relevant angles, one immediately obtains $I_{X/Y} = S\pm S_x$ and $I_{A/D} = S\pm S_y$.
Substitution of the numerically computed periodic time dependence $\mathbf{S}(t)$ with the use of the above relations in the definition of the second-order correlation function \eqref{eq:g2} readily reproduces its oscillations with the decay time $\tau$.
Damping of these oscillations is, in turn, governed by fluctuations affecting the condensate density, as the condensate global phase along with its fluctuations cancels out in $g^{(2)}(\tau,\mathbf{r}_1,\mathbf{r}_2)$ by definition \eqref{eq:g2}.
Such fluctuations due to either fast reservoir kinetics or slow pump intensity noise are described with a stochastic term in Eqs.\eqref{eq:resdyn}, moreover, it is minimally sufficient to assume stochastic time variation of the axially symmetric pump part $W$.


\begin{figure}
    \centering
    \includegraphics[width=\linewidth]{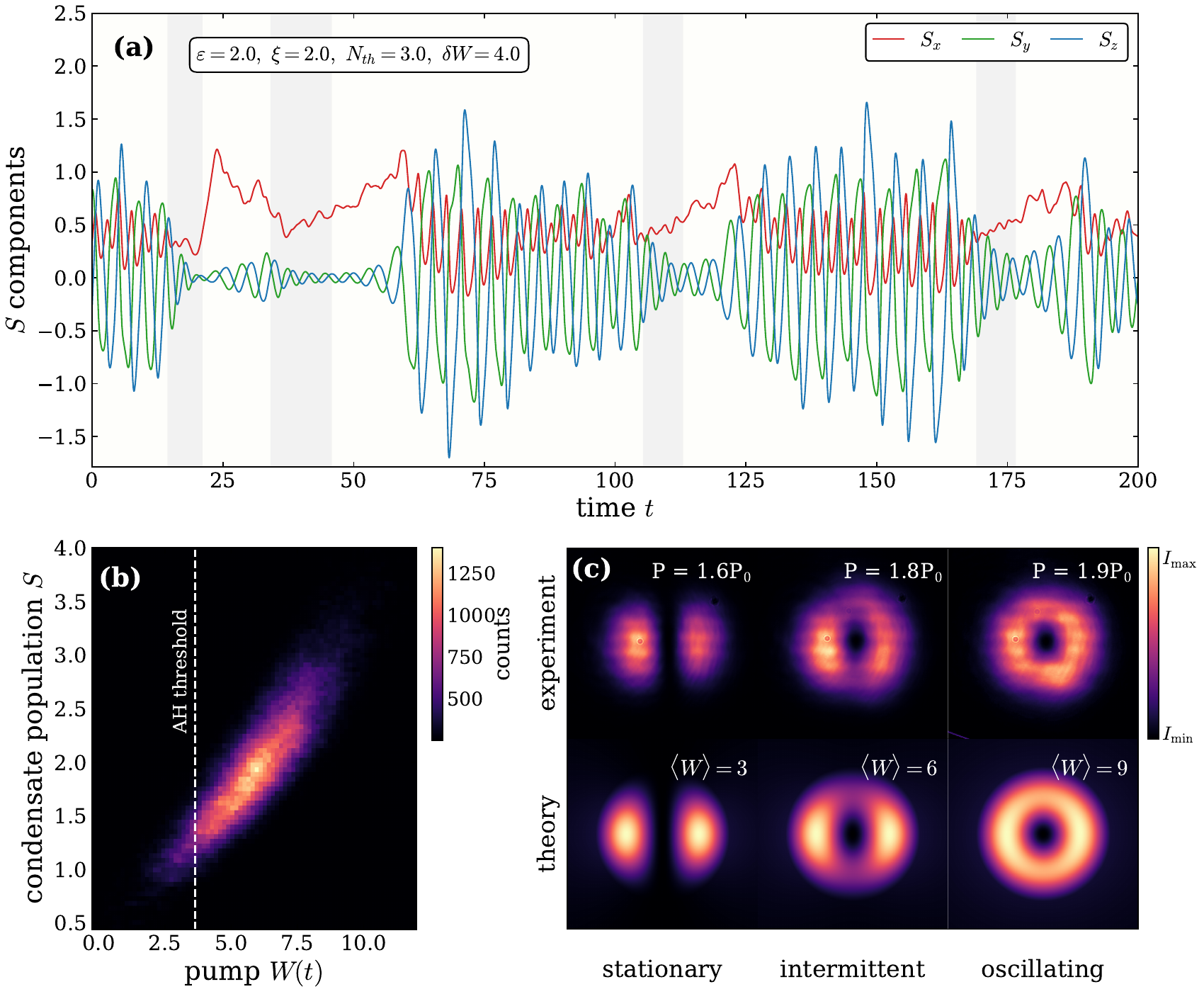}
    \caption{Intermittent condensate dynamics.
    (a) Stochastic switching between quasi-stationary below-bifurcation (light blue) and oscillating above-threshold (light yellow) regimes visible in the time dependence of the pseudovector components $S_x,S_y,S_z$.
    (b) The histogram of the pump power $W$ and condensate population $S$ relative to the AH bifurcation (dashed vertical line).
    (c) Time-averaged polariton condensate density profiles, illustrating a crossover from the quasi-stationary mode to the oscillating regime through the intermittent phase with increasing mean pump power in experiment (top row) and in theory (bottom row).}
    \label{fig:four}
\end{figure}

Fig. \ref{fig:two}f shows the numerically computed oscillations of $g^{(2)}(\tau)$ for the four experimental configurations in the oscillating regime at mean pump power $\langle W \rangle_t = 12$ well above the bifurcation threshold $W_\text{AH} \approx 3.6$.
High oscillation amplitudes of $g^{(2)}_{A,D}$ and $g^{(2)}_{A,D}$ contrasting weak oscillations of $g^{(2)}_{X,Y}$ and $g^{(2)}_{Y,Y}$ reflect large ratio of the $S_y$ and $S_x$ amplitudes, in qualitative agreement with the high-intensity experimental data in Fig.\ref{fig:two}с.
At lower mean pump powers $\langle W \rangle_t$ near the AH bifurcation, the system enters the intermittent dynamics regime, in which fluctuations induce stochastic transitions between the stationary and oscillating condensate modes.
These transitions significantly affect the measured $g^{(2)}(\tau)$ profiles (Fig.\ref{fig:two}a), which are reproduced with slow pump fluctuations defined by low mean pump power $\langle W \rangle_t = 6$, standard deviation $\sigma_W=1.7$, and frequency cut-off $f_c = 0.2$ (Fig.\ref{fig:two}c).

Fig. \ref{fig:four} shows the intermittent condensate dynamics regime in which fluctuations induce stochastic transitions between the two qualitatively different modes.
These transitions are driven by pump power $W(t)$ crossing the AH bifurcation threshold $W_\text{AH}$ at timescales that are slow compared to both the oscillation period and the characteristic condensate-reservoir response times, as shown in Figs. \ref{fig:four}a.
Notably, even a relatively small net probability of finding the condensate below the bifurcation threshold, shown in Fig.\ref{fig:four}b, results in strongly deviating from unity $g^{(2)}(\tau)$ for all four considered two-point configurations.
This highlights the sensitivity of our experimental approach based on measuring second-order correlations to bifurcations and its capacity to reveal stochastic intermittent dynamics.
In contrast, time-averaged condensate density profiles in the deterministic oscillating and stochastic intermittent regimes can be virtually indistinguishable, as shown in Fig. \ref{fig:four}c.
Furthermore, the donut-shaped polariton density profiles such as in our experiments can also be attributed to stationary polariton vortex states \cite{Alyatkin2024}, thus time-resolved measurements are essential for revealing the true dynamics of optically trapped polariton condensates.


The two-point spatiotemporal correlation function $g^{(2)}(\tau,\mathbf{r}_1,\mathbf{r}_2)$ oscillates over extremely long times $\sim2\,$ns, exceeding both polariton lifetime $\sim10\,$ps and condensate coherence time $\sim350\,$ps, revealing persistent periodic condensate dynamics.
The correlation measurements allowed us to unravel the details of this oscillatory regime by mapping it onto pseudospin precession around an effective field controlled by the trap ellipticity.
This essentially linear behaviour, similar to quantum beats, is unexpected in an open interacting many-body system and is, paradoxically, governed by its nonlinear dynamics.
Specifically, a combination of repulsive condensate self-interaction and reservoir-mediated effective attraction leads to soft dynamical instability of a trivial fixed-point attractor, corresponding to a condensate formed at the highest-gain eigenstate of the complex non-Hermitian trapping potential \cite{Nalitov2019,Chestnov2024}.
In this scenario, corresponding to a supercritical AH bifurcation, the destabilized fixed-point attractor transforms into a stable limit cycle where the condensate pseudospin is persistently precessing over a closed path on the Bloch sphere.
Most notably, this limit cycle is uniquely stable in the regime of partial mutual compensation of repulsive and attractive types of nonlinearity, where its periodicity is determined by the energy splitting of eigenstates, governed by the Schr\"{o}edinger equation, in agreement with previously reported results \cite{opt6040053,Barrat2024}.

The experimental method implemented here fully reveals its advantages at the pump power slightly exceeding the condensation threshold where we have been able to resolve the presence of the AH bifurcation in asymmetric distortion of the second-order correlation function spatiotemporal profile.
Near the bifurcation pumping threshold, fluctuation-driven transitions between the fixed-point and limit-cycle modes in the intermittent regime give rise to significant deviations from unity in two-point configurations, in stark contrast to the deterministic limit-cycle mode.
These transient processes, occuring in stochastic dynamics of bistable or multistable polariton condensates \cite{Aleiner2012,Ohadi2015,Barrat2024}, are lost in the time-averaged emission data.

The authors thank F.\,P.\,Laussy and I.\,V.\,Ignatiev for fruitful discussions.
A.K. acknowledges the Saint Petersburg State University for the Research Grant No. 125022803069-4.
A.N. acknowledges support from the Ministry of Science and Higher Education of the Russian Federation (Goszadaniye) (No. FSMG-2026-0012).

\bibliography{Library} 
\end{document}